\documentclass{pasj00}

\begin{document}
\SetRunningHead{Shimizu et al.}{Hinode SOT and XRT Coalignment Calibration}
\Received{2000/12/31}
\Accepted{2001/01/01}

\title{Hinode Calibration for Precise Image Co-alignment between SOT and XRT
(November 2006 -- April 2007)}


\author{%
  Toshifumi \textsc{Shimizu}\altaffilmark{1},
  Yukio \textsc{Katsukawa}\altaffilmark{2},
  Keiichi \textsc{Matsuzaki}\altaffilmark{1},
  Kiyoshi \textsc{Ichimoto}\altaffilmark{2},
  Ryohei \textsc{Kano}\altaffilmark{2},
  Edward E. \textsc{DeLuca}\altaffilmark{3},
  Loraine L. \textsc{Lundquist}\altaffilmark{3},
  Mark A. \textsc{Weber}\altaffilmark{3},
  Theodore D. \textsc{Tarbell}\altaffilmark{4},
  Richard A. \textsc{Shine}\altaffilmark{4},
  Mitsuru \textsc{S\^oma}\altaffilmark{2},
  Saku \textsc{Tsuneta}\altaffilmark{2},
  Taro \textsc{Sakao}\altaffilmark{1},
  and
  Kenji \textsc{Minesugi}\altaffilmark{1}
}
\altaffiltext{1}{Institute of Space and Astronautical Science (ISAS),
 Japan Aerospace Exploration Agency (JAXA), \\
 3-1-1 Yoshinodai, Sagamihara, Kanagawa, 229-8510}
\email{shimizu.toshifumi@isas.jaxa.jp}
\altaffiltext{2}{National Astronomical Observatory of Japan,
 Mitaka, Tokyo 181-8588}
\altaffiltext{3}{Harvard-Smithsonian Center for Astrophysics, Cambridge,
 MA 02138, U.S.A.}
\altaffiltext{4}{Lockheed Martin Solar and Astrophysics Laboratory,
 Bldg. 252, 3251 Hanover St., Palo Alto, CA 94304, U.S.A.}




\maketitle

\begin{abstract}
To understand the physical mechanisms for activity
and heating in the solar atmosphere, the magnetic coupling
from the photosphere to the corona is an important piece of
information from the Hinode observations, and therefore
precise positional alignment is required among
the data acquired by different telescopes.
The Hinode spacecraft and
its onboard telescopes were developed to allow us to investigate
magnetic coupling with co-alignment accuracy better than 1 arcsec.
Using the Mercury transit observed on 8 November 2006 and co-alignment
measurements regularly performed on a weekly basis, we have determined
the information necessary for precise image co-alignment and
have confirmed that co-alignment better than 1 arcsec can be
realized between Solar Optical Telescope (SOT) and X-Ray Telescope (XRT)
with our baseline co-alignment method. This paper presents results from
the calibration for precise co-alignment of CCD images from SOT and XRT.

\end{abstract}

\section{Introduction}

Many of the observational studies by the Hinode spacecraft to address 
magnetic couplings in the solar atmosphere need to combine the data from 
some of the three telescopes onboard the spacecraft (Kosugi et al.\ 2007).
After its successful launch
and its early operation, the Solar Optical Telescope (SOT, 
Tsuneta et al. \ 2007; Tarbell et al.\ 2007; Suematsu et al.\ 2007; 
Shimizu et al.\ 2007; Ichimoto et al.\ 2007)
has started to produce series of 0.2--0.3 arcsec
visible-light images, giving the dynamical behavior of solar magnetic
fields on the solar surface. Simultaneously, the X-ray Telescope (XRT,
Golub et al.; Kano et al.\ 2007) has been providing 1 arcsec resolution
X-ray images of the solar corona, giving the location of heating and
dynamics occurring in the corona.
Precise image co-alignment of SOT and XRT data with sub-arcsec
accuracy (0.5 arcsec as our goal) is required
to provide new information
regarding magnetic couplings from the photosphere to the corona.

The satellite structure was designed with careful consideration to
meet the telescope alignment requirements: the 1 hour stability must be
5 arcsec (0-p), and the static DC offset in the pointing of each telescope
must be 80 arcsec (0-p), to allow telescopes with narrow fields of view to
observe the same observing target. Here the static DC offset means 
the static deviation on the pointing direction of one telescope from 
that of the other telescope on orbit after experiencing the launch 
mechanical environment. The sub-arcsec accuracy required on
the image co-alignment is realized with calibration using
scientific images acquired during the flight.
The three telescopes are aligned along the Z axis of the spacecraft and
supported by an optical bench unit (OBU). The OBU is a cylinder made
up of composite material that internally supports the optical telescope
part (OTA; Suematsu et al.\ 2007) of the SOT.
The SOT's focal plane package (FPP; Tarbell et al.\ 2007),
EIS (EUV imaging spectrometer, Culhane et al. 2007) and XRT are
kinematically mounted on the outside of the OBU
with 6 mounting legs each, which constrain the degrees of freedom of
the rigid body. The OBU also holds a tower to whose upper surface
the sun sensors are attached.

\begin{figure*}[t]
  \begin{center}
    \FigureFile(120mm,130mm){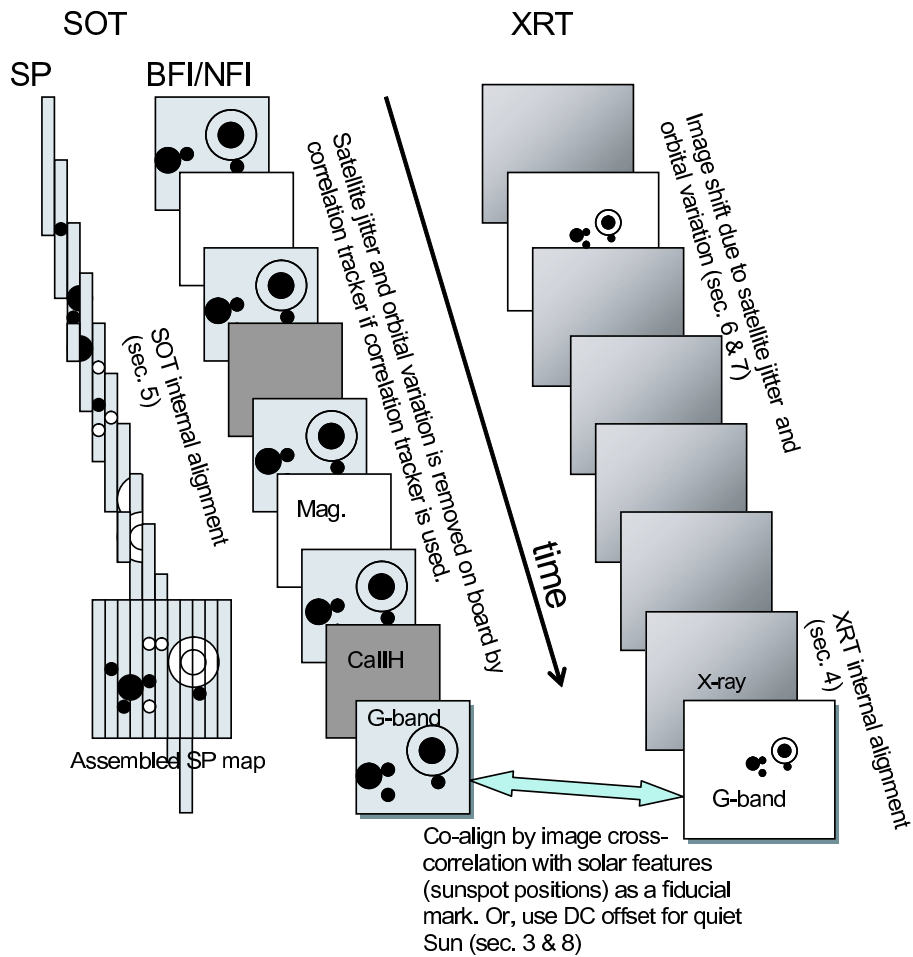}
  \end{center}
  \caption{Schematic explanation of the baseline method for the
    most accurate co-alignment of the Hinode data.}
  \label{fig: concept}
\end{figure*}

For achieving sub-arcsec accuracy in co-aligning the images to each other,
a baseline co-alignment concept adapted to the Hinode
data was defined in the early phase of the spacecraft development,
which is briefly summarized in section \ref{sec: method}.
After each telescope started initial observations in late October
2006, the DC pointing offsets among the telescopes were determined
with the data acquired during the Mercury transit on the solar disk.
This provides us with a unique calibration opportunity; we are able
to accurately determine the DC pointing offset, the plate scale, and
the roll angle of the CCD images from solar north (section \ref{sec: mercury}). The XRT can take both G-band images,
through the Visible Light Imager, and X-ray images; the offsets
between these two imaging systems are discussed in section \ref{sec: xrt}.
Similarly for SOT, images taken through different filters are
slightly offset from each other, as discussed in section \ref{sec: sot}.
The pointing direction of each telescope slightly varies with time, primarily
associated with the orbital phase of the satellite. Section
\ref{sec: orbit} shows how the pointing direction of each telescope
and sun sensors changes with orbital phase. This information is used
when the time series of XRT images is co-aligned.
Section \ref{sec: correction} describes the method using
the orbital variation model and sun sensor signals to remove
the satellite jitter and longer time drift due to orbital variation.
This alignment method is shown to give excellent performance, i.e.,
residual jitter of 0.3 arcsec at the 1 $\sigma$ level or better.
Finally, section \ref{sec: absolute} shows how the DC offset between
SOT and XRT pointing varies from November 2006 to April 2007. This
information can allow us to establish a more general method
for co-alignment between SOT and XRT if we correctly manage
the behavior of the tip-tilt mirror in SOT. The general method
is crucial for quiet Sun studies.

This paper does not discuss the co-alignment of SOT and XRT with data from EIS, which observes EUV spectral lines with slit or slot
scanning of an observing region for diagnosing
the properties of the plasma in the transition region and lower
corona. This is because we need more calibration efforts to register
the assembled scans of EIS spectral data to the rigid CCD frames
from XRT and SOT. The co-alignment of SOT and XRT with EIS data will 
be presented
in a separate paper after completing the calibration efforts.

\section{SOT/XRT co-alignment method}
\label{sec: method}

The baseline method for SOT/XRT co-alignment of Hinode
data is schematically described in Figure \ref{fig: concept}.
Co-aligning the time series of the images from each telescope needs
to be performed before co-aligning SOT images with XRT images.
The XRT images need to be corrected for both high frequency spacecraft
jitter (1--2 arcsec p-p) and orbital distortions associated with
the thermal deformations (a few arcsec p-p). We show that the jitter
can be removed by applying corrections based on the sun sensor signals
with an empirical model for the orbital
distortions (section \ref{sec: correction}).
In contrast, the SOT correlation tracker (Shimizu et al.\ 2007) removes
most of the satellite jitter and orbital variation from the series of
SOT images. However, there is a slow drift originating from
the gross motion of solar granules seen in the narrow ($11 \times 11$ arcsec)
field of view of the correlation tracker, which can be removed by
performing a rigid co-alignment using cross correlation from image to
image and then applying the cumulative offsets to the whole time series.

To obtain the best alignment, XRT and SOT need to take regular G-band
images. If the images are taken every 10 minutes then co-alignment of
less than 1 arcsec can be attained.
When sunspots or pores are located in the field of view, they are used
as fiducials to co-align the data from the two telescopes with each other.
In quiet sun data, there are no clear fiducial marks in 1 arcsec G-band
images from XRT, and the calibrated DC pointing offset needs to be used for
co-alignment (section \ref{sec: absolute}).

SOT can take images at different
wavelengths with its filter imaging capabilities. The Broadband Filter Imager
(BFI) produces photometric images with broad spectral resolution in
6 bands at the highest spatial resolution and at rapid cadence.
The Narrowband Filter Imager (NFI) provides intensity, Doppler, and
full Stokes polarimetric imaging at high spatial resolution.
Slight offsets and different magnifications exist between the images
at different wavelengths, but they can be co-aligned with
calibrated offset and magnification parameters
(section \ref{sec: sot}). SOT has another observing capability,
the Spectro-Polarimeter (SP), which obtains line profiles of two magnetically
sensitive Fe lines and the nearby continuum while slit scanning a region of
interest. The solar features will evolve and move during a fairly
long scanning time, but a small portion of the scanning area can be well
co-aligned with BFI or NFI images by using the granule patterns and
other solar features seen in the continuum.


\section{Mercury transit and DC pointing offset}
\label{sec: mercury}


\begin{figure}
  \begin{center}
    \FigureFile(80mm,120mm){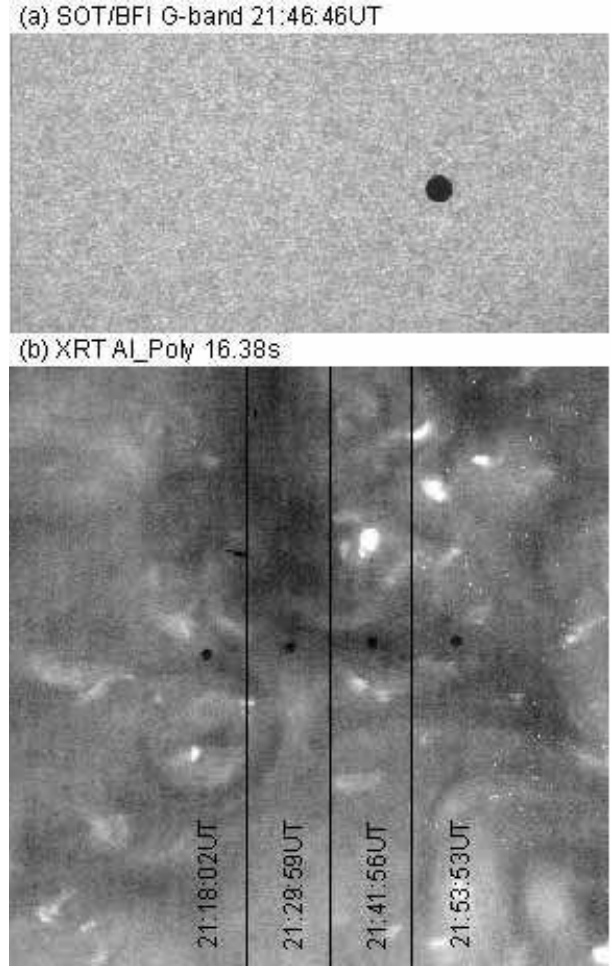}
  \end{center}
  \caption{Mercury transit observed with (a) SOT and (b) XRT on
     8 November 2006. The small dark spot in each frame is Mercury.}
  \label{fig: mercury}
\end{figure}

An unique opportunity for obtaining calibration data came about 10 days
after opening the protective doors of the telescopes. A transit of Mercury
across the solar disk was observed on 8 November 2006.
The images acquired in 21:18 - 21:55 UT were used for this calibration.
Mercury was seen in 5 successive frames of G-band exposures
($4K \times 2K$ pixels, no pixel summing) by SOT BFI. One of the images
taken with SOT is shown in Figure \ref{fig: mercury} (a).
XRT observed the Mercury transit with a lot of X-ray exposures
($512 \times 512$ pixels, no pixel summing, Al\_Poly filter, 16.38 sec
exposure duration), four frames of which are shown in
Figure \ref{fig: mercury} (b) to give the movement of Mercury in this
period. During this period, only series of X-ray exposures were made, at
different focus positions, for evaluating the best focus position using
the sharp limb of Mercury.

\begin{figure}
  \begin{center}
    \FigureFile(82mm,180mm){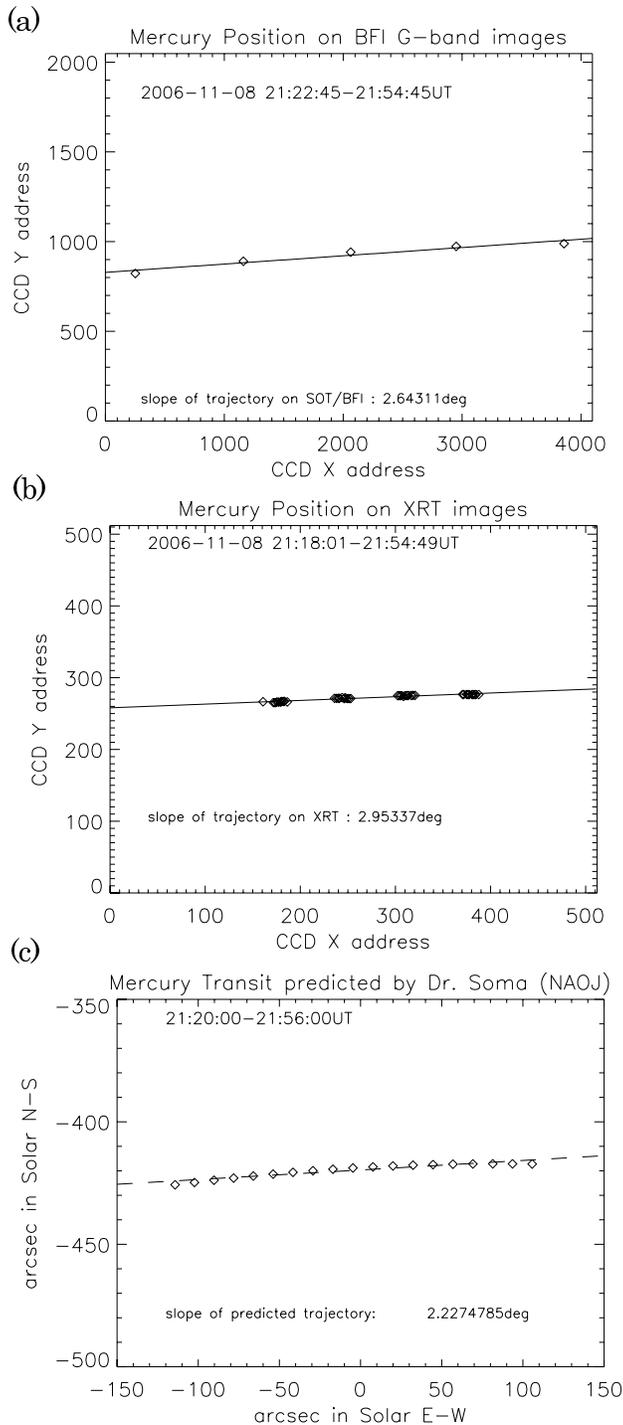}
  \end{center}
  \caption{Trajectory of Mercury on the CCDs: (a) Position of Mercury
    in SOT BFI G-band $2K \times 4K$ CCD frames. (b) Position of
    Mercury observed with XRT X-ray Al\_Poly filter $512 \times 512$ pixel
    frames.
    The CCD positions after performing positional corrections to remove
    satellite jitter and orbital variation for XRT and tip-tilt mirror
    angle for SOT. (c) Prediction of Mercury position in the solar
    heliocentric coordinate seen from the Hinode satellite.}
  \label{fig: trajectory}
\end{figure}

\begin{figure}
  \begin{center}
    \FigureFile(82mm,90mm){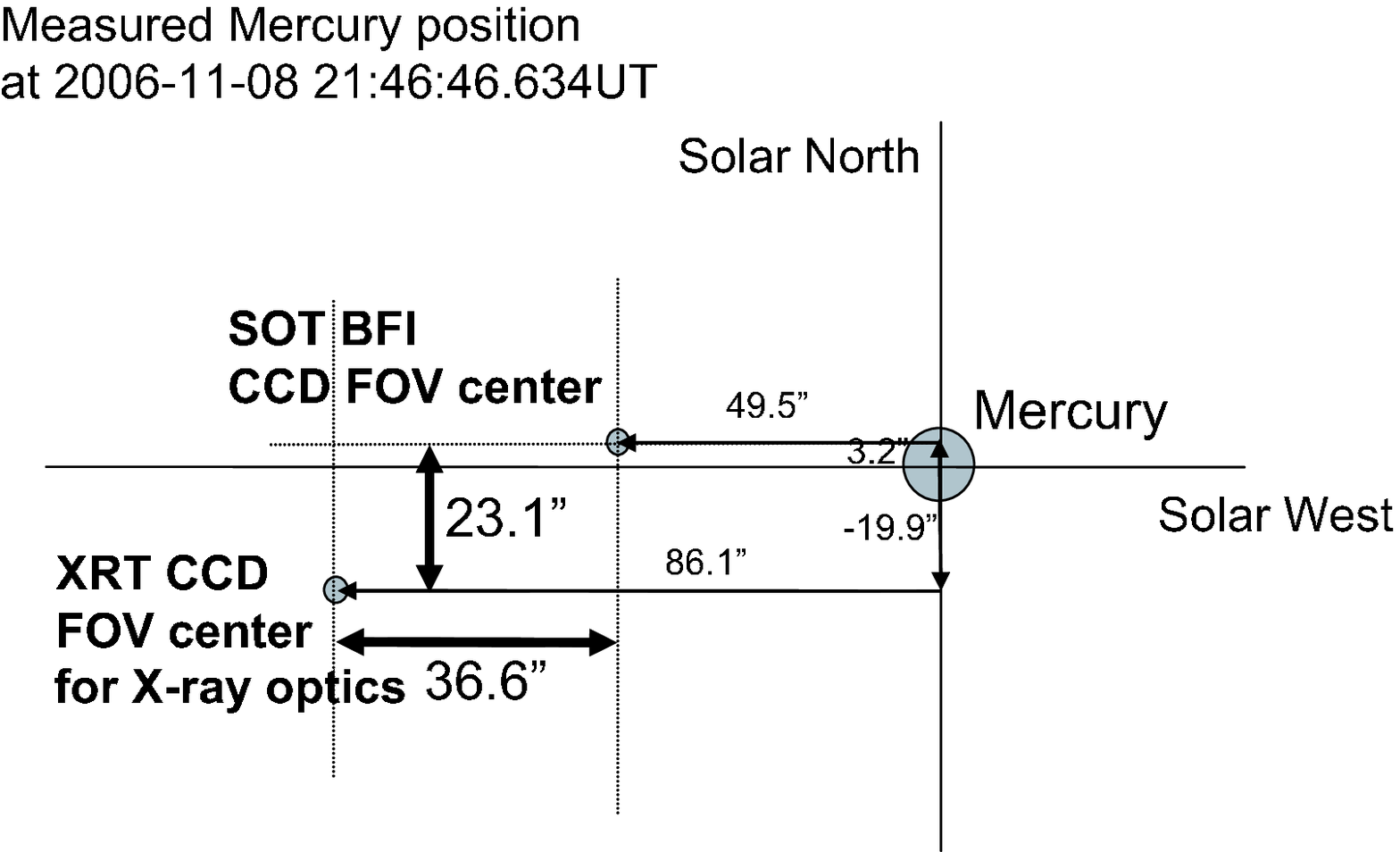}
  \end{center}
  \caption{DC pointing offset between SOT and XRT X-ray images.}
  \label{fig: dcoffset}
\end{figure}

The position of Mercury can be used to measure the offset between the SOT
field of view and the XRT X-ray field of view. Since the size of Mercury
is small (10.0 arcsec in diameter), the position of Mercury
in each of frames was determined with sub-arcsec accuracy, as shown
in Figure \ref{fig: trajectory} (a) and (b). Note that the time stamp
in UT was verified by using the timing of Mercury touching
to each of the solar limbs. Satellite jitter as well as orbital variation
were removed in XRT data according to the information described in section
\ref{sec: correction}. Active tip-tilt mirror control with correlation
tracker (Shimizu et al.\ 2007) was enabled during the period. At the time
between the 3rd and 4th frames, the tip-tilt mirror angle was reset
to its home position (note that it is slightly biased due to hysteresis,
see Shimizu et al.\ 2007 for details) because the angle reached
the stroke limit, causing a positional jump in the series of
SOT images. The magnitude of the positional jump was determined
and corrected (4.9 arcsec W, 0.0 arcsec N) by using magnetic patterns that 
were well observed in Ca II H frames, and that were also acquired in G-band.
Figure \ref{fig: dcoffset} gives the pointing offset between the XRT X-ray
optics and SOT BFI, as determined with the Mercury transit.
The center of the XRT X-ray field of view
is located 36.6 arcsec east and 23.1 arcsec south of the BFI CCD
field of view.

The transit of Mercury also gives the plate scale and orientation of the CCD
frames by comparing the measured trajectory on the CCD frames with
the prediction of the Mercury transit in solar heliocentric coordinate.
It is noted that a similar calibration with another Mercury transit
was made for the soft X-ray images from the {\em Yohkoh} satellite
(W\"ulser et al.\ 1998). Figure \ref{fig: trajectory} (c) is
the position of Mercury on the solar disk calculated using orbital
elements of the Hinode spacecraft.
The plate scale derived from the Mercury transit is as follows:

 $0.0545 \pm 0.0004$ arcsec/pixel for SOT BFI CCD;

 $1.031 \pm 0.001$ arcsec/pixel for XRT CCD.

Note that a small uncertainty is included in the measurement for the BFI CCD
because of imperfect tip-tilt mirror angle correction.
The derived XRT plate scale is in good agreement with the plate scale derived
from the fitting to the X-ray limb,
which gives $1.031 \pm 0.001$ arcsec pixel$^{-1}$.
The data taken on 21 December 2006 at 11:00--14:00 UT gives $951.0 \pm 1.0$
pixel for the radius of the solar dark disk determined by the limb fitting,
whereas the radius of the visible-light photospheric sun is 975.45 arcsec
at that date. It was also assumed that the radius of the X-ray dark disk is
3500 km larger than that of the visible-light photospheric disk with
consideration of the distance from the photosphere to the bottom of
the corona. Moreover, another precise comparison can be made to confirm
the XRT plate scale (Ishibashi 2007)--- the limb position of an XRT
X-ray image was co-aligned with the limb of an Fe {\sc xv} 284\AA\ image
from {\em SoHO} EIT (Delaboudini\`ere et al.\ 1996), which plate
scale has been well-calibrated. This comparison gives an XRT plate scale 
statistically consistent with that from the Mercury measurement.

\begin{figure}
  \begin{center}
    \FigureFile(80mm,120mm){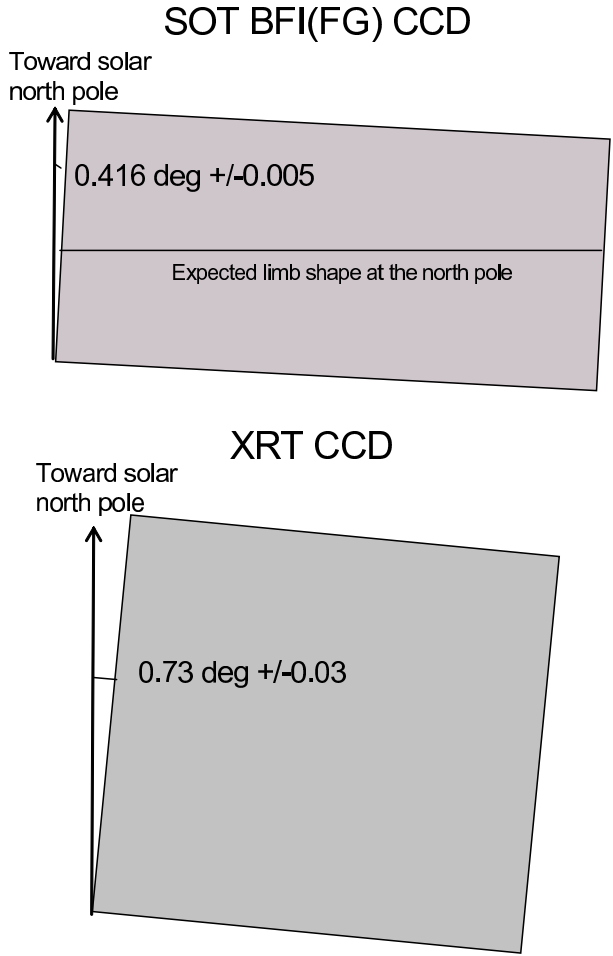}
  \end{center}
  \caption{Roll angle offset of
    SOT/BFI and XRT CCD frames from the solar north direction.}
  \label{fig: roll}
\end{figure}

The slope of trajectory observed on each CCD gives the roll angle offset
of the frames from the solar north direction. The prediction gives
that the slope of the trajectory in the period of 21:20 -- 21:55 UT
is tilted 2.228 deg from the east-west line. The slope of the Mercury
trajectory is tilted 2.643 deg on the BFI CCD frame and is tilted 2.95 deg
on the XRT CCD frame. This measurement gives that
the BFI CCD frame is offset $0.416 \pm 0.005$ deg clockwise,
and the XRT CCD frame is offset $0.73 \pm 0.03$ deg clockwise, both
from the solar north direction (Figure \ref{fig: roll}).

\section{XRT internal offset}
\label{sec: xrt}

The XRT consists of optimized Wolter-I-like grazing incidence
soft X-ray optics together with co-focal visible light optics
(Golub et al.\ 2007). Both soft X-ray and visible light images are
focused onto a single CCD detector (Kano et al.\ 2007).
G-band images are taken by inserting the glass G-band
filter into the focal plane and opening the visible light shutter.
This allows light passing through the visible light optic
to reach the CCD, and the glass filter is opaque to X-rays.
The G-band optical axis has a slight offset from the X-ray optics, although
the magnification (plate scale) is same.
The offset was measured before launch at the X-Ray Calibration Facility
(XRCF) at NASA/MSFC (NASA Marshall Space Flight Center) using an optical 
light source (Golub et al.\ 2007).
The design of the XRT mirror support plate ensured that
the visible light optic and the X-ray mirror remained rigidly
fixed through launch.

\begin{figure}
  \begin{center}
    \FigureFile(70mm,70mm){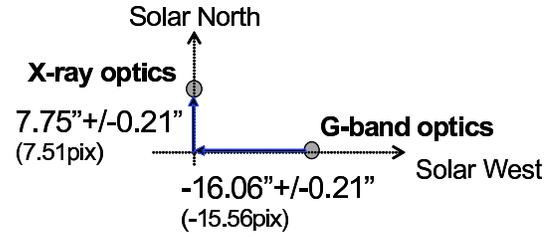}
  \end{center}
  \caption{XRT internal offset of X-ray images from G-band images.}
  \label{fig: xrtoffset}
\end{figure}

On orbit we have checked the offset by
fitting the solar limb seen in both G-band and X-ray images
to determine both the center of the Sun and the radius in
CCD pixels.
The series of G-band and X-ray images acquired on
3 March 2007 at 6:05--9:30 UT (22 frames for each) were used.
Since it is known that the limb fitting routine used is sensitive to the
initial guess, the initial guess was selected to be located roughly at the middle
of the travel range due to orbital variation. It was also checked that the
initial guess does not change the results even when $\pm 1$ pixel
perturbation for (x, y, r) is applied. All the results were matched
within $<$ 0.07 pixel.
Short-term satellite jitter and orbital drift corrections
were applied to the fitting results.
The offset of XRT G-band images from X-ray images determined by this study is given
in Figure \ref{fig: xrtoffset}.
The error of each G-band/X-ray pair is 0.8--0.9 pixel, but 22 pairs of measurements
reduce the error to 0.2 arcsec.
The alignment between the G-band and X-ray optics is stable; no systematic
offset change is associated with orbital phase.

\section{SOT internal offset}
\label{sec: sot}

Slight offset and pixel scale (magnification) differences exist
from image to image at different wavelengths.
They were evaluated with 6 sets of the images taken at synoptic observations
in November 2006 -- April 2007, by performing a rigid co-alignment using cross
correlation from image to image.
Tables \ref{tab:bfi} and \ref{tab:nfi} give the derived shift
in X (E-W) direction and Y (N-S) direction, and the derived 
scale deviations, in
comparison to BFI G-band images (4305 \AA).
Table \ref{tab:bfi} shows that, for example, the field-of-view (FOV) center of
Ca II H (3968\AA) full (4K $\times$ 2K) frames is located 1.35 pixels
east and 5.24 pixels south of the FOV center in G-band full frames,
where the pixel unit is the pixel of original Ca II H images before
scaling its magnification.
Table \ref{tab:nfi} gives that the FOV center of Fe I line (6302\AA)
full frames is 3.4 pixels east and 26.3 pixels north of
the FOV center in G-band full frames, where the pixel unit is
the pixel of the Fe I images before scaling its magnification.
Since NFI has not yet acquired
a sufficient number of images at the other wavelengths, the pre-launch
evaluation (Okamoto et al.\ 2007) is given in Table \ref{tab:nfi}.
The accuracies of the derived offset and scale differences are estimated to
be about 0.2 pixels at the 1$\sigma$ level and 0.00005, respectively.

\begin{table}
  \caption{BFI Images offset and scale.}\label{tab:bfi}
  \begin{center}
    \begin{tabular}{lllll}
      \hline
      wavelength & xshift$^{(1)}$ & yshift$^{(1)}$ & xscale$^{(2)}$
        & yscale$^{(2)}$ \\
      (\AA)     & (pix)  & (pix) &         &        \\
      \hline
      3883 & 0.94 & -6.23 & 1.00041 & 1.00041 \\
      3968 & -1.35 & -5.24 & 1.00021 & 1.00021 \\
      4305 & 0.00 & 0.00 & 1.00000 & 1.00000 \\
      4504 & 1.34 & 6.49 & 0.99921 & 0.99921 \\
      5550 & -0.36 & 3.03 & 0.99497 & 0.99497 \\
      6684 & 0.60 & -1.38 & 0.99140 & 0.99140 \\
      \hline
    \end{tabular}
  \end{center}
   {\small
      Note (1): The offset at the center pixel (2047.5, 1023.5) of
       the full frame (4K$\times$2K) images to the center pixel of
       G-band (4305) data. The offset is given in the pixel unit
       of original image at
       each wavelength before scaling its magnification. \\
      Note (2): Scale deviation from the G-band data. The value
       larger than 1 means that the pixel scale of original image
       at each wavelength is larger than that of G-band image.
   }
\end{table}

\begin{table}
  \caption{NFI Images offset and scale in reference of G-band (4305)
     center.}\label{tab:nfi}
  \begin{center}
    \begin{tabular}{lllll}
      \hline
      wavelength & xshift$^{(1)}$ & yshift$^{(1)}$ &
      xscale$^{(2)}$ & yscale$^{(2)}$ \\
      (\AA)     & (pix)  & (pix) &         &        \\
      \hline
       5172* & -6.2 & 23.9 & 1.4720 & 1.4650 \\
       5250* & -6.6 & 23.3 & 1.4711 & 1.4642 \\
       5576* & -6.2 & 24.5 & 1.4699 & 1.4630 \\
       5896* & -8.2 & 22.1 & 1.4673 & 1.4604 \\
       6302  & -3.4 & 26.3 & 1.4660 & 1.4591 \\
       6563* & -2.8 & 25.9 & 1.4655 & 1.4586 \\
      \hline
    \end{tabular}
  \end{center}
  {\small
       Note (1) (2): same as notes in Table 1. \\
      Note *: Not yet evaluated with flight data. Pre-launch data
       (Okamoto et al.\ 2007) is shown for reference only.
  }
\end{table}

\begin{figure*}[t]
  \begin{center}
    \FigureFile(160mm,160mm){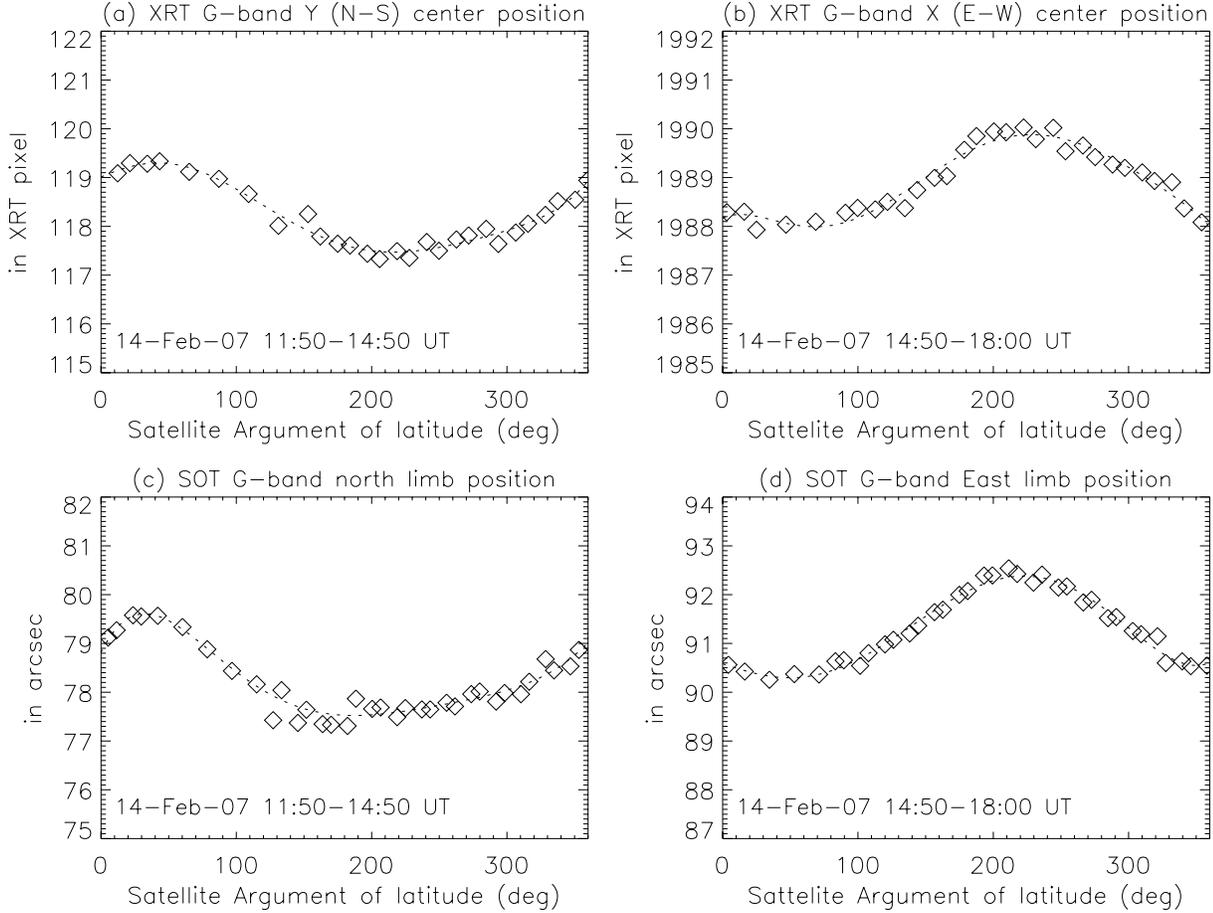}
  \end{center}
  \caption{SOT and XRT pointing direction as a function of orbital phase.}
  \label{fig: orbital}
\end{figure*}

The spectro-polarimeter (SP) data needs a careful consideration when
it is co-aligned with the BFI and NFI images. Each slit position
may contain a small deviation from the ideal scanning step
(0.16 arcsec), as shown in Okamoto et al.\ (2007).
Also, the alignment of SP optics is significantly
sensitive to ambient temperature, and the calibrated level-1 data
may still contain positional jitter on the order of about 1 arcsec
along the slit,
although the calibration performs a correction for the alignment change
due to ambient temperature. Moreover, solar features evolve and
move during the relatively long period of each map. With these
characteristics, accurate co-alignment of SP data with BFI and NFI images
would be achieved only when a rigid co-alignment
using cross-correlation is applied to a limited
scanning range of the SP data.

\begin{figure*}[bt]
  \begin{center}
    \FigureFile(140mm,60mm){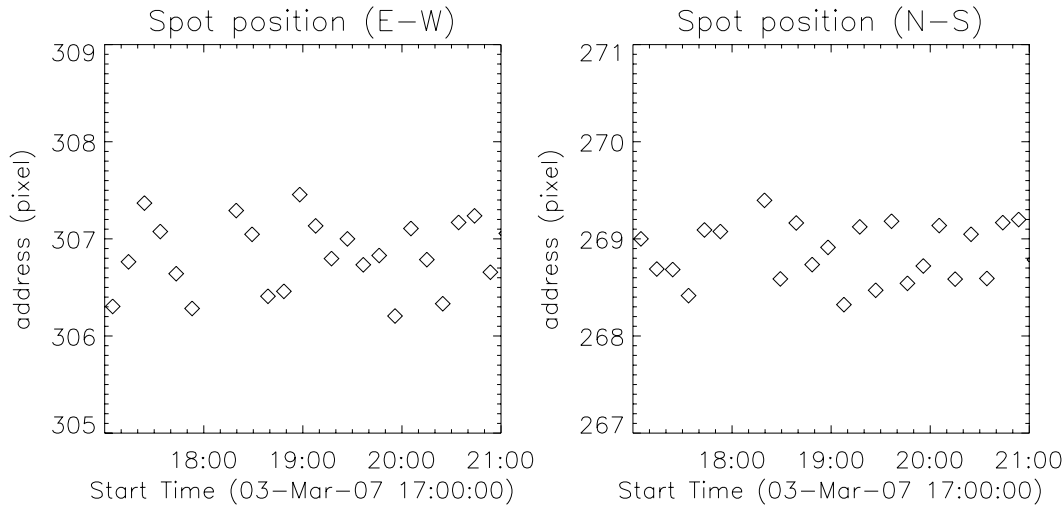}
  \end{center}
  \caption{Sunspot position as a function of time after applying the satellite
    jitter and orbital drift corrections. The center position of
     the leading sunspot of NOAA 10944 is derived from the time series
     of G-band images acquired on 3 March 2007. The pixel unit is 1.031 
     arcsec.}
  \label{fig: xrtjitter}
\end{figure*}

\section{Pointing drift associated with orbital phase}
\label{sec: orbit}

Each telescope onboard the Hinode spacecraft shows pointing
drift on the order of a few arcsec (p-p), which varies as
a function of the orbital phase of the satellite. In order
to monitor the orbital variation regularly, the team has performed weekly-basis co-alignment measurements since February 2007.
These measurements are limb
observations to the north and the east, with about 2 hours runtime for each of the limbs.
Using the limb seen in the CCD frames from each telescope, we can
monitor how the pointing of each telescope behaves with time.

Figure \ref{fig: orbital} is the orbital drift of XRT and SOT fields of view
as a function of satellite latitude, measured on 14 February 2007.
Disk center position on the CCD frames is derived from a series of XRT
G-band images by limb fitting, whereas the position of the partial portion
of the limb seen in each SOT frame is monitored with a series of BFI G-band
images. The correlation tracker is disabled, and the tip-tilt mirror is 
at the home
position with accuracy $<$ 0.3 arcsec during the measurements.
Pointing jitter induced by the satellite is removed by subtracting sun
sensor signal, and therefore the alignment offset of each telescope
from one of the sun sensors as the reference is derived. It turned out
that the orbital drift of the XRT pointing is quite similar to that of
the SOT pointing. The orbital drift is well repeated with
the period of the satellite revolution.

\section{Satellite jitter and orbital drift correction}
\label{sec: correction}

Signals from two sun sensors, Ultra Fine Sun Sensors UFSS-A and UFSS-B,
can be used to know the jitter induced by the satellite attitude body
control. The signals have good accuracy with a random noise level of
0.35 arcsec at the 3$\sigma$ level. 
It should be noted that the UFSS sensor signals also
contain a gradual orbital drift that depends upon
the orbital phase of the satellite.
We have established a method (xrt\_jitter.pro) to
apply a satellite jitter and orbital drift correction to
a time series of XRT images by using the sun sensor signals and
the information described in section \ref{sec: orbit}.
As an example, Figure \ref{fig: xrtjitter} shows the jitter residual in
the time series of XRT G-band images (acquired over about 4 hours)
after applying the correction to the image cube. Sunspot position seen
in the field of view is given as a function of time.
The series of images are stabilized within residual jitter of 1 arcsec
(p-p) or better.

\begin{figure}
  \begin{center}
    \FigureFile(80mm,160mm){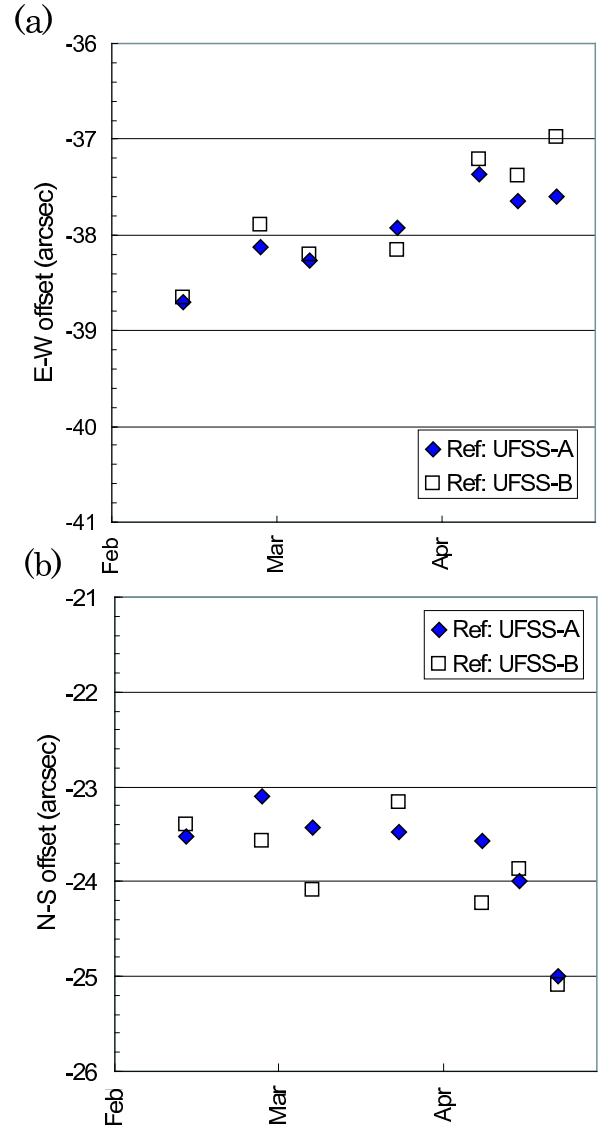}
  \end{center}
  \caption{Long-term change of the pointing offset between SOT and XRT.}
  \label{fig: trend}
\end{figure}

\section{Long-term evolution of the DC pointing offset}
\label{sec: absolute}

According to the baseline co-alignment method in section \ref{sec: method},
the offset between the XRT and SOT data needs to be 
determined for each dataset.
This is because we do not have the information on the tip-tilt mirror
angle with sufficient accuracy for subarcsec co-alignment. This method can
work very well for active-region observations, in which the position of
sunspots in G-band data can be utilized to derive the offset. However,
the DC pointing offset value is needed for co-aligning the images of
the quiet Sun, which have no fiducial in G-band, although the co-alignment
accuracy may be degraded due to uncertainty in the tip-tilt mirror angle.
Moreover, XRT did not acquire G-band images regularly during observations
at the performance verification phase before early February, so that
we need the pointing offset even for some active-region observations.

Since February 2007, co-alignment measurements give how the DC pointing
offset changes over long time scales, 
which is shown in Figure \ref{fig: trend}.
This shows that the offset between XRT and SOT is extremely rigid but that
it has slowly drifted about 2 arcsec in both N-S and E-W directions in
about 3 months. Note that the absolute values in the figure may contain
1--2 arcsec uncertainty due to systematic error in plate scale, because
the center position of the solar disk in the BFI frame coordinate is estimated
from the measured position of the partial limb seen in the narrow BFI field
of view using the solar radius from the calendar, which is then
compared with the center of the solar disk derived from XRT fitted
limbs. The comparison of the sunspot position observed both with BFI and
XRT G-band data at almost the same time (12 UT on 28 February) gives
that the offset in N-S and E-W directions is 24.2 arcsec and
37.9 arcsec, respectively.

After 26 February 2007, a special tip-tilt mirror reset is implemented
at the start of correlation tracker operation (CT SERVO ON in observation
time line) to minimize the bias caused by hysteresis of the tip-tilt
mirror mechanism. This reset
ensures that the tip-tilt mirror starts from its home
position angle within 0.1 arcsec at the 1$\sigma$ level.


\section{Final remarks}

We have evaluated the co-alignment performance of Hinode observational
data, especially the co-alignment between SOT and XRT data, and
confirmed that co-alignment better than 1 arcsec can be realized
between SOT and XRT with our baseline method. The results
presented in this paper can be applied to the data acquired before early
May 2007, the start of the eclipse season (May -- August).
The co-alignment performance will be changed significantly during
the eclipse season, and the DC pointing offset may be changed
before and after the eclipse season. Hinode data users
should remember this note.

\vspace*{12pt}

Hinode is a Japanese mission developed and launched by ISAS/JAXA, with NAOJ
as domestic partner and NASA and STFC (UK) as international partners. It is
operated by these agencies in co-operation with ESA and NSC (Norway).
The authors would like to express their thanks to all the people who have
been involved in design, development, tests, launch operation, and science
operations for realizing the Hinode (Solar-B) mission and its new
advanced observations presented in this paper. For archiving sub-arcsec
co-alignment among the data from the onboard telescopes as one of key
technical requirements on the spacecraft design, spacecraft system
engineers of Mitsubishi Electric Corp., especially Sadanori Shimada,
Toshio Inoue, Norimasa Yoshida, are greatly acknowledged for their
efforts on the spacecraft structural and thermal designs.


\end{document}